\documentclass[fleqn,10pt]{wlscirep}
\usepackage[utf8]{inputenc}
\usepackage[T1]{fontenc}
\usepackage{cite}
\title{Complementary Speckle Stimulated Emission Depletion Microscopy}

\author[1,*]{Payvand Arjmand}
\author[2]{Samlan Chandran Thodika}
\author[1]{Haoyang Li}
\author[1]{Elsa Bivas}
\author[1]{Martin Oheim}
\author[3]{Hiroyuki Yoshida}
\author[2]{Etienne Brasselet}
\author[1,4,5,*]{Marc Guillon}

\affil[1]{Saints-Peres Paris Institute for the Neurosciences, CNRS, Universite Paris Cite, Paris, 75006, France}
\affil[2]{Laboratoire Ondes et Matière d'Aquitaine, CNRS, Universite de Bordeaux, Talence, F-33400, France}
\affil[3]{School of Engineering, Kwansei Gakuin University, Sanda, Hyogo 669-1330, Japan}
\affil[4]{Institut Langevin, ESPCI Paris, PSL University, CNRS, 75005, Paris, France}
\affil[5]{Institut Universitaire de France, 75007, Paris, France}

\affil[*]{payvand.arjmand@polytechnique.edu}
\affil[*]{marc.guillon@u-paris.fr}

\begin{abstract}
Stimulated emission depletion (STED) microscopy has
emerged as a powerful technique providing visualization of biological
structures at the molecular level in living samples. In this technique, the
diffraction limit is broken by selectively depleting the fluorophore’s excited
state by stimulated emission, typically using a donut-shaped optical vortex
beam. STED microscopy performs exceptionally well in degraded optical
conditions, such as living tissues. Nevertheless, photobleaching and
acquisition time are among the main challenges for imaging large volumetric
fields of view. In this regard, random light beams such as speckle patterns
have proved to be especially promising for three-dimensional imaging in
compressed sensing schemes. Taking advantage of the high spatial density of
intrinsic optical vortices in speckles-one of the most commonly used types
of structured beams in STED microscopy-we propose here a novel scheme
that employs speckles for performing STED microscopy. Two speckle patterns are generated at the excitation and the depletion
wavelengths, respectively, exhibiting inverted intensity contrasts. We illustrate spatial resolution enhancement using complementary
speckles as excitation and depletion beams on both fluorescent beads and biological samples. Our results establish a robust method
for super-resolved three-dimensional imaging with promising perspectives in terms of temporal resolution and photobleaching.\par
\vspace{5mm}
\end{abstract}

\begin{document}

\flushbottom
\maketitle
%
%
\thispagestyle{empty}


\section*{Introduction}
Super-resolution microscopy has significantly enhanced our ability to visualize biological structures with nanoscopic details. Stochastic approaches like single-molecule localization microscopy (SMLM)\cite{betzig2006imaging,rust2006stochastic} have become very popular in cell biology due to their simplicity. However, the sequential localization of individual fluorescent molecules and their sensitivity to aberrations makes them unsuited to image fast living systems.\\
In contrast, targeted approaches based on patterned laser illumination like structured illumination microscopy~\cite{gustafsson2005nonlinear} and stimulated emission depletion microscopy (STED) \cite{hell1994breaking, klar2000fluorescence} can be made fast by reducing the field of view~\cite{movement2001video} or by parallelizing detection~ \cite{yang2014large, bergermann20152000}. Furthermore, the optical singularity used to break the diffraction limit is especially robust to optical aberrations. The fast acquisition rate together with the robustness to aberrations makes it especially efficient to image dynamic processes in living cells \cite{westphal2008video, schneider2015ultrafast}, tissues\cite{ding2009supraresolution, berning2012nanoscopy} and even in living animals~\cite{berning2012nanoscopy}. As a limitation, under intense laser illumination, the full sample volume is illuminated while only the signal from the focus region is collected, which induces unnecessary photo-toxicity for the sample~\cite{oracz2017photobleaching} and compromises the possibility of imaging the volume. Several techniques implement different strategies to either avoid unnecessary exposure of fluorophores to the depletion laser or protect them from the damaging effects~\cite{danzl2016coordinate, gottfert2017strong, staudt2011far}. The recent so-called MINFLUX~\cite{balzarotti2017nanometer} and MINSTED~\cite{weber2021minsted} approaches super-localize individual molecules by triangulation using respectively an excitation and depletion donut-shaped intensity profiles as two distinct means to minimize the amount of required photons. So far, these methods have been limited to sparse labels, being highly sensitive to spurious background signals. Sample-adaptive scanning approaches \cite{heine2017adaptive, dreier2019smart, alvelid2022event} have achieved real-time adjustments to acquisition conditions to capture data at a reasonable rate, simultaneously minimizing photobleaching, processing time, and data volume. Here we propose an efficient use of all emitted photons from thick biological samples together with the increase of the volumetric field of view while maintaining the high-resolution and the high-speed capabilities of STED microscopy.
\begin{figure}[h!]
	\centering
	\includegraphics[width=\linewidth]{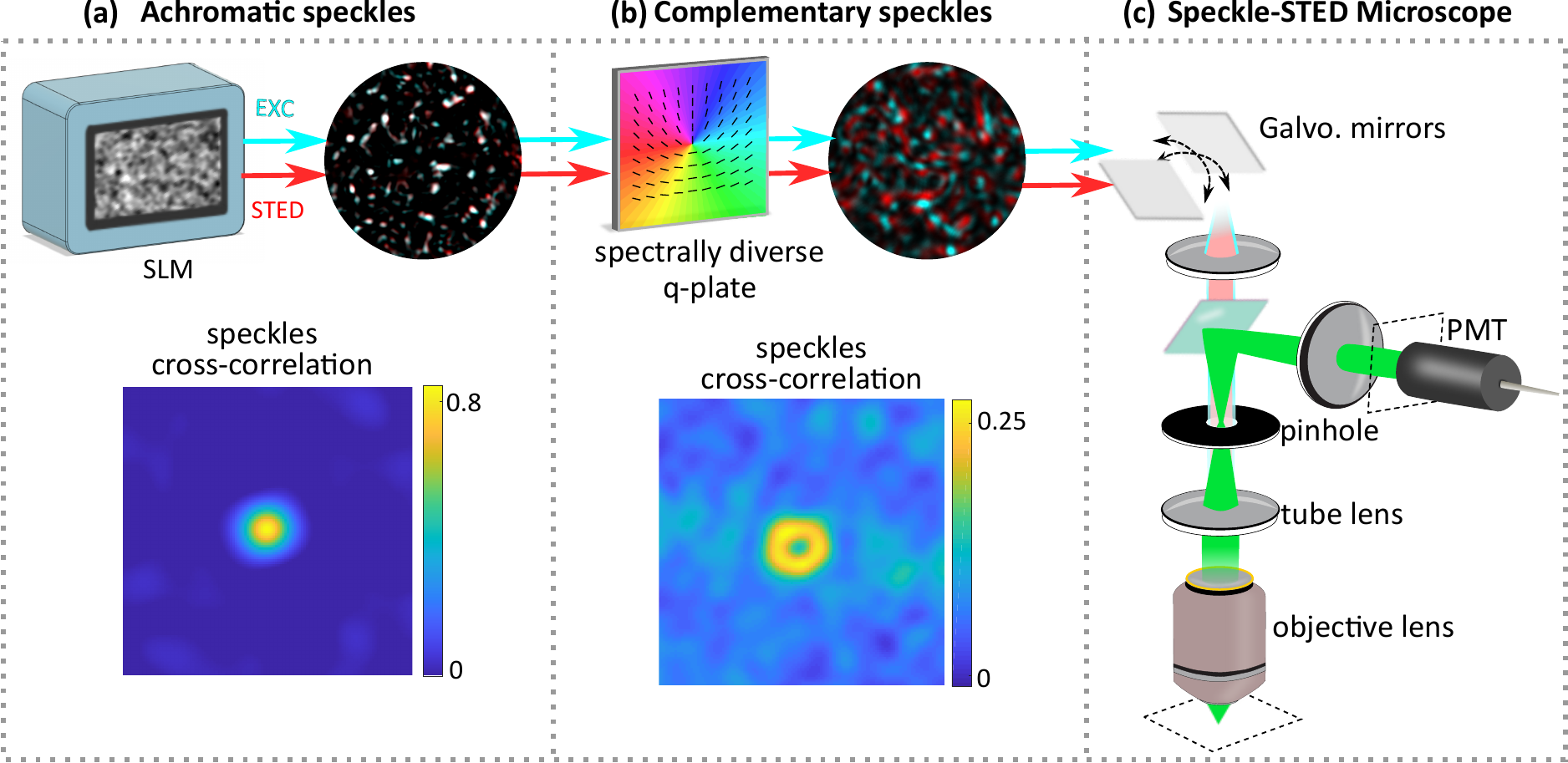}
	\caption{The principle of complementary speckle-STED microscopy. (a) The excitation and STED linearly polarized pulsed laser beams at $\lambda_{EXC}=639 nm$ and $\lambda_{STED}=775 nm$, with 1 MHz repetition rate, are combined before the SLM using a dichroic mirror. A random smooth Gaussian phase mask is projected on the SLM to generate highly correlated speckle patterns, as illustrated by the cross-correlation product $\lambda_{EXC}\star\lambda_{STED}$. (b) A so-called spectrally diverse q-plate is placed in the Fourier-conjugated plane with respect to the plane of the SLM, which acts as a vortex phase modulator at $\lambda_{STED}$ and as a flat phase mask at $\lambda_{EXC}$, having prepared the polarization state of the two incident fields as co-circular using an achromatic quarter waveplate. This allows generating two complementary speckle patterns with complementary random intensity patterns.  (c) A pair of galvanometric mirrors are then used to scan the sample. The fluorescent signal is epi-detected through the same objective and separated from the illumination path through a dichroic mirror. The signal is collected by a hybrid photomultiplier tube.}
	\label{fig:principle}
\end{figure}
\\Our approach relies on combining  recent advances in optical instrumentation and computational imaging. In structured illumination microscopy , structured patterns such as fringes\cite{gustafsson2000surpassing}, grids\cite{chakrova2015studying}, or speckles\cite{mudry2012structured,yilmaz2015speckle,yeh2017structured} are employed to get both optical sectioning and resolution improvement over large field of views. The diffraction barrier can be further broken by saturating optical transitions~\cite{gustafsson2005nonlinear, fujita2007high, pascucci2016superresolution, bender2021circumventing}.  
In this context, we recently proposed a speckle scanning microscope providing super-resolution imaging in 3D from a single 2D raster scan only. The diffraction barrier was broken by saturating fluorescence excitation~\cite{pascucci2019compressive}. In this technique, the whole photon budget emitted in a volume of interest of the sample (and collected by the microscope objective) is recorded. Optical sectioning is achieved using a large confocal pinhole and three-dimensional resolution in the optically-sectioned volume of interest is enabled thanks to the randomly structured illumination 3D-pattern~\cite{lim2008wide} . Random wavefields such as speckle patterns exhibit key characteristics that make them ideally suited for compressed sensing schemes\cite{donoho2006compressed} . Speckles appearing in axial planes distant by more than the Rayleigh range are statistically orthogonal~\cite{bernet2011lensless}, so enabling three dimensional imaging from a single two-dimensional image only~\cite{antipa2018diffusercam, pascucci2019compressive}. The previous solution we proposed was essentially limited by background noise generated by the intense excitation pulses required for saturating fluorescence. 
Here, in contrast to this previous speckle-based compressive imaging scheme we propose to saturate the stimulated emission rather than the fluorescence excitation, which allows getting rid of drawbacks from sample luminescence, since there is virtually no emission associated with the intense beam for the considered wavelengths. In addition, red-shifting the super-resolving intense laser reduces photo-bleaching and photo-toxicity, and provides enhanced optical penetration depth.\par
In STED microscopy, two beams with distinct wavelengths are employed, one exciting fluorescence while the other induces stimulated emission. Structuring of the latter laser line into a optical vortex beam exhibiting a intensity zero allows shrinking the size of the fluorescence-emitting region. Optimal resolution and signal to noise ratios are obtained when the excitation maximum coincide with the zero of the de-exciting beam. The latter principle applies to light fields embedding multiple zeros of intensity~\cite{bergermann20152000}, and shall thus be further extended to the case of speckle patterns that contain dense networks of optical vortices with topological charges $+1$ and $-1$~\cite{nye1974dislocations}.
Indeed, optical vortices in speckle patterns can confine light to subdiffraction dimensions~\cite{pascucci2016superresolution}. Tailoring their spatial distribution statistics can further improve the fluorescence confinement~\cite{bender2018customizing}.
Said differently, a speckle pattern can play the role of the multi-site depletion beam to saturate an optical transition similarly to the case of multi-donut structures used in parallelized RESOLFT microscopy~\cite{chmyrov2013nanoscopy}.\par

While uniform excitation illumination could simplify the setup and still allow higher excitation
power,as done in parallelized STED microscopy~\cite{yang2014large,bergermann20152000}, our simulations (see Supporting Section 1) suggest that it would result in lower imaging resolution than complementary
speckle illumination and also lower contrast. Additionally, the speckle pattern
approach offers advantages in robustness to aberrations, scattering, and interferences, which
are beneficial for achieving enhanced spatial resolution. Yet it remains challenging to implement speckle patterns for excitation and depletion experimentally, especially for two different wavelengths.
For a given wavelength, a spiral phase mask used as a Fourier filter switches the location of intensity maxima and optical vortices according to a cyclic permutation algebra~\cite{gateau2017complementary,gateau2019topological}. This scheme would be directly implementable for the case of single-wavelength two-photon STED microscopy~\cite{bianchini2012single}. However, extending the same approach to two different wavelengths implies getting control of random interferences in presence of optical path length chromatic dispersion that tends to shorten the spectral correlation widths. In this context, it has been recently unveiled that forward scattering samples such as biological tissues do generate highly achromatic speckle patterns~\cite{vesga2019focusing,zhu2020chromato,arjmand2021three}. Here we report on the combination of the generation of two-wavelength achromatic speckle field with vortex Fourier filtering speckle manipulation, which implies wavelength selective shaping of optical vortices.\\

The principle of our method is illustrated in Fig.~\ref{fig:principle}. First, a two-wavelength achromatic speckle is generated using a spatial light modulator (SLM) for our two laser lines at (639nm and 775nm wavelengths Fig~\ref{fig:principle}a). Then we use a so-called spectrally diverse q-plate ~\cite{yan2015q} placed in the Fourier-conjugated plane with respect to the plane of the SLM, which acts as a vortex phase modulator at $\lambda_{STED}$ and as a flat phase mask at $\lambda_{EXC}$. Owing to a Fourier lens, two complementary speckle patterns are thus obtained and further conjugated with the sample plane (Fig.~\ref{fig:principle}b), the latter being imaged using a speckle-STED microscope ~\cite{pascucci2019compressive} that is basically a confocal microscope with a large detection pinhole to collect fluorescence light over a large (and tunable) confocal volume (Fig.~\ref{fig:principle}c).\\ 
\section*{Experimental design}
\begin{figure}[ht!]
	\centering
	\includegraphics[width=\linewidth]{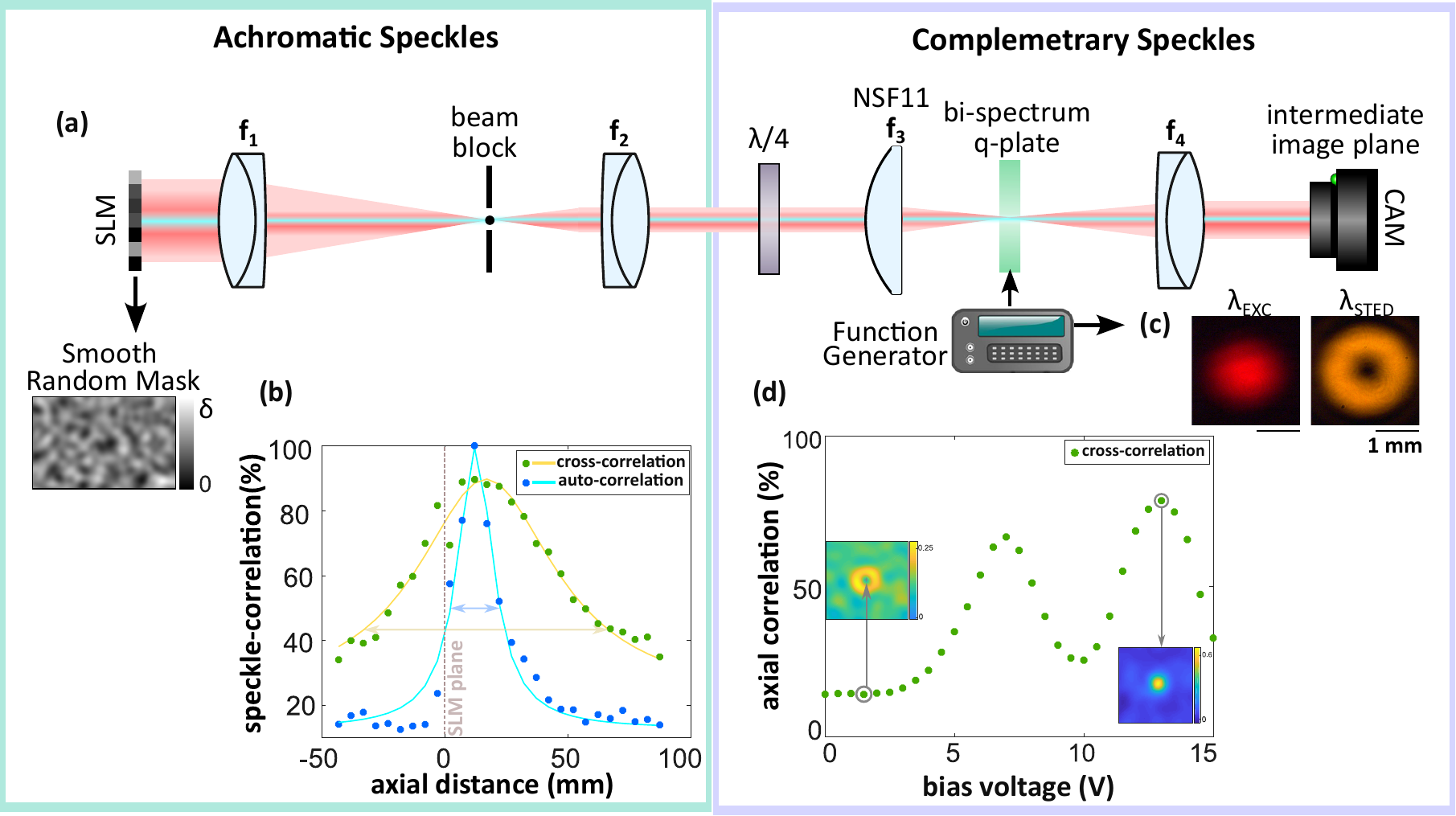}
	\caption{Optimization of optical parameters of the diffuser mask (right box) and of the bi-spectrum bi-modal mask (left box).  (a)  Experimental setups for achromatic and complementary speckle generation. The first telescope $f_1 = 300$ mm and $f_2 = 30$ mm is used to reduce the size of the beam. The beam block is placed at the Fourier plane of $f_1$ . The second telescope is $f_3 = 50$ mm and $f_4 = 100$ mm to image the SLM on an intermediate image plane on a camera. $f_3$ is a plano-convex lens of dispersive material NSF11 to compensate for the chromato-axial shift of the two beams. The q-plate is placed at the Fourier plane of $f_3$. (b) Experimental cross-correlation product between the excitation and STED speckle intensity patterns vs the propagation distance (green markers). Solid curve: Lorentzian fit. (c) Experimental auto-correlation of the excitation beam with the speckle pattern of the maximum correlation plane (blue markers). Solid curve: Lorentzian fit . (c) Gaussian beams being unmodulated and spiral-phase modulated by the bi-spectrum q-plate at the excitation and the STED wavelengths, respectively.
		(d) Cross-correlation between excitation and STED speckle intensity patterns \emph{vs.} applied voltage on the liquid crystal cell.}
	\label{fig:achro-compl}
\end{figure}

Although forward scattering optical elements generate "achromatic" speckle patterns ~\cite{vesga2019focusing,zhu2020chromato,arjmand2021three}, we opt for the use of a spatial light modulation that allows controlling the characteristics of the speckle field. Optical phase modulation is performed in the limit of small modulation depth, for which the pure phase transmittance mask can be expanded as $e^{ik\delta}\simeq 1 + ik\delta$, where $k=2\pi/\lambda$ is the wavevector and $\delta(x,y)$ is the spatial distribution of the optical path delay imparted by the SLM to the incident light. Up to a scaling factor $k$ on the intensity, this ensures the achromatic nature of the scattered intensity pattern once undiffracted light is blocked, as long as the first-order Taylor expansion holds. In practice, the $\delta(x,y)$ profile is obtained by filtering a random pixelated phase map with values uniformly spreading over a $0-2\pi$ range by a Gaussian kernel:  
\begin{equation*}
	G(r)=\frac{G_0}{2\pi w^2}\exp\left[\frac{-r^2}{2w^2}\right]
\end{equation*}
as described in Ref.~\cite{guillon2017vortex} (Fig.~\ref{fig:achro-compl}a, green box). The random phase diffuser is thus characterized by two parameters: the correlation width $w$ and the standard deviation of the optical path delay $\overline{\delta} = G_0/(2w)$.
Here, the achromatic behavior of the diffuser is obtained at the expense of an increase in the fraction of undiffracted light that is proportional to $exp\left[-(k\overline{\delta})^2\right]$ and that is removed thanks to a filter beam block in a Fourier plane (Fig.~\ref{fig:achro-compl}a). The main drawback of this approach is the loss in laser power at the sample plane. 
To minimize this loss, a compromise must be found between the amount of undiffracted light energy and the achromaticity of the generated speckles, which can be met by playing on the standard deviation of $\overline{\delta}$ (See Supporting Section 2). 
As a result, two-wavelength smooth and achromatic random light fields $\propto \delta(x,y)$ can be obtained on-demand in an image plane of the SLM. However these smooth-phase random wavefields do not exhibit any optical vortex -- required for performing super-resolution -- in the conjugate plane of the SLM. For the random intensity statistics to be fully developed, free space propagation over a sufficient distance is required.
For the two patterns to remain achromatic after propagation a relay telescope is used wherein one of the two lens is made of the dispersive material N-SF11 (Fig.~\ref{fig:achro-compl}a). The achromatic speckle plane is then axially shifted at a distance where they are fully developed and contain optical vortices. The maximum spatial correlation for the two speckle fields is found at a plane at a distance $z=15~{\rm mm}$ from the image plane of the SLM (represented as $z=0$ in Fig.~\ref{fig:achro-compl}b). The correlation between the speckle patterns at this plane is $88\%$ (Fig.\ref{fig:achro-compl}.b). We calculate the full-width half-maximum (FWHM) ${\rm d}z$ by fitting our experimental data with the expected Lorentzian function~\cite{zhu2020chromato,arjmand2021three} defined as $L=\alpha/[(z-z_0)^2+\beta]$. 
The axial extent of the two-wavelength speckle correlation must be compared to the one of speckle grains, the ratio between the two axial extents qualitatively indicating the maximum number of planes that can be imaged. We measured that the FWHM of the speckle axial auto-correlation at wavelength $\lambda_{EXC}$ is a factor $3.72$ smaller than the achromatic speckle axial range so allowing simultaneous STED imaging of $~4$ planes. 
\\The optical element providing complementary speckle patterns for the excitation and STED fields can be described as an inhomogeneous and anisotropic waveplate performing the following operation on an incident circularly polarized paraxial light field propagating along the z axis, up to an unimportant phase factor:
\begin{equation}
	\textbf{c}_{+} \rightarrow   \cos(\Delta/2) \textbf{c}_{+} + i \sin(\Delta/2) e^{\pm i\phi} \textbf{c}_{\mp}
\end{equation}
where $\textbf{c}_{\pm}=(x\pm iy)/\sqrt{2}$ refers to the unit circular polarization Jones vector and $\Delta=2\pi(n_e-n_o)d/\lambda$ refers to the birefringent phase retardation, with $d$ the thickness of the anisotropic slab, and $n_e$ and $n_o$ the extraordinary and ordinary refractive indices of the anisotropic material whose optical axis orientation angle in the plane $(x,y)$ is $\psi=\phi/2$, where $\phi$ is the usual polar angle. This allows imparting a spiraling phase profile of the form $exp(\pm i \phi)$ to the STED field while leaving the excitation field unaffected by appropriately designing the thickness of the optical element and the anisotropy of the chosen material.
In practice, we prepared a liquid crystal q-plate whose extraordinary refractive index can be adjusted electrically, similarly to previous works~\cite{piccirillo2010photon}, made of 5CB nematic liquid crystal (from Merck).  By adjusting the applied voltage on the liquid crystal cell, the extraordinary refractive index can be varied from its maximal value (1.6977 at $\lambda_{EXC}$ and 1.6822 at $\lambda_{STED}$ at room temperature~\cite{tkachenko2006nematic}) at zero voltage to its minimal value equal to the ordinary refractive index (1.5298 at $\lambda_{EXC}$ and 1.5225 at $\lambda_{STED}$ at 25°C temperature~\cite{tkachenko2006nematic}) at high voltage. The thickness of the liquid crystal plate is chosen to be large enough to ensure that the two retardances at the two wavelengths of interest ideally satisfy $\Delta(\lambda_{EXC})=2n\pi$ and  $\Delta(\lambda_{STED})=(2m-1)\pi$ with $n$ and $m$ integers. Neglecting the dispersion of the optical anisotropy, the previous criterion is rewritten as : 
\begin{equation}
	\frac{\lambda_{EXC}}{\lambda_{STED}} = \frac{2n}{2m-1}
\end{equation}
Since $\lambda_{EXC}/\lambda_{STED} =1.213\approx6/5$, Eq.(2) is compatible with the choice n=m=3, which is found to be realized experimentally using $d=9$ $\mu m$ under an applied voltage $U=0.930$ $V_{rms}$ at 1kHz frequency. Fig.~\ref{fig:achro-compl}c shows the total intensity profiles at the output of the device for an input Gaussian beam with diameter of about $1.5$ $mm$, for both the excitation and the STED wavelengths while Fig.~\ref{fig:achro-compl}d equivalently shows the speckle cross-correlation maps.\\ 
We would like to emphasize that our technique for generating speckle patterns is different from imaging techniques relying on dynamic speckle illumination that typically require several random realizations ($\sim$1000). The reason why our approach does not require so many speckle grains is due to the “speckle spot congruence”~\cite{freund19981001} describing that all speckle spots in a speckle pattern are very similar in shape. Therefore, the auto-correlation is very isotropic even when averaging on a small number of speckle grains. 
\section*{Results}
We implemented our two-wavelength complementary speckle engineering system onto a beam scanning microscope~(Supporting Section 3). Two linearly polarized pulsed lasers at 1 MHz are used: an excitation laser at $\lambda_{exc}=639~{\rm nm}$ (PDL 800-D, Picoquant) and a STED laser at $\lambda_{STED}=775~{\rm nm}$ (Katana 08-HP, NKT Photonics). The two beams are synchronized using a digital pulse delay generator (T560, Highland Technology), combined using a dichroic beam-splitter and reflected on the SLM where the previously characterized smooth random Gaussian phase mask is projected to generate achromatic and complementary speckle patterns. A pair of galvanometric mirrors are used to scan the sample. To optimize the axial component of the field at vortex locations, a quarter-wave plate is positioned just before the objective (Olympus UPLanSApo, 100x, NA = 1.4) to change the polarization from linear to circular to ensure isotropic resolution improvement~\cite{pascucci2016superresolution, pascucci2019compressive,bender2021circumventing}. The fluorescent signal is epi-detected through the same objective and separated from the illumination path through a dichroic beam-splitter. The signal is then detected by a hybrid photomultiplier tubefluorescence through a bandpass filter (ET706/95m,
Chroma). The electronic signal is then acquired using a fast digitizer (PCIe EON Express 12-Bit ENE-123-G20, GaGe by Vitrek, Lockport, IL, USA).\par
In order to characterize the fluorescence depletion as a function of power of the STED beam and tune the delay between the excitation and the depletion pulses, the two beams are first prepared in a same speckle pattern mode. For this characterization, the achromatic speckles are obtained by tuning the driving voltage of the bi-spectrum q-plate at $V = 13 V$ (Fig.~\ref{fig:achro-compl}d). The plane where the correlation between the two speckles is maximum (the achromatic plane) is conjugated to the focal plane of the microscope objective. 
One isolated fluorescent $200~{\rm nm}$ nanobead is then scanned in the transverse plane while gradually increasing the STED beam power. According to theory, the normalized average fluorescence signal originating from two achromatic fully developed speckle illumination -- for excitation and saturated stimulated emission depletion -- is expected to vary as (Supporting Section 4):
\begin{equation}
	\overline{I_F} = \overline{I_{E}}\left (\frac{1}{1+s/\sigma}\right)^2
\end{equation}
where $s = \overline{I_{STED}}/I_{sat}$, 
the saturation parameter, is the ratio between the average speckle intensity of the depletion beam and the saturation intensity of the stimulated emission process and $\sigma$ is a correction factor to adjust for experimental measurement errors. The saturation of fluorescence depletion as a function of the STED beam power shown in Supporting Section 4, demonstrates efficient depletion for the (estimated) maximum pulse energy of $2~{\rm \mu J}$ provided by our laser at the sample, over a speckle extending over a disk-shaped support of $4~{\rm \mu m}$ in diameter. Note that, even though the speckles are not spatially homogeneous, the image rendering is homogenized by the speckle scanning principle.\par
\begin{figure}[h!]
	\centering
	\includegraphics[scale=0.6]{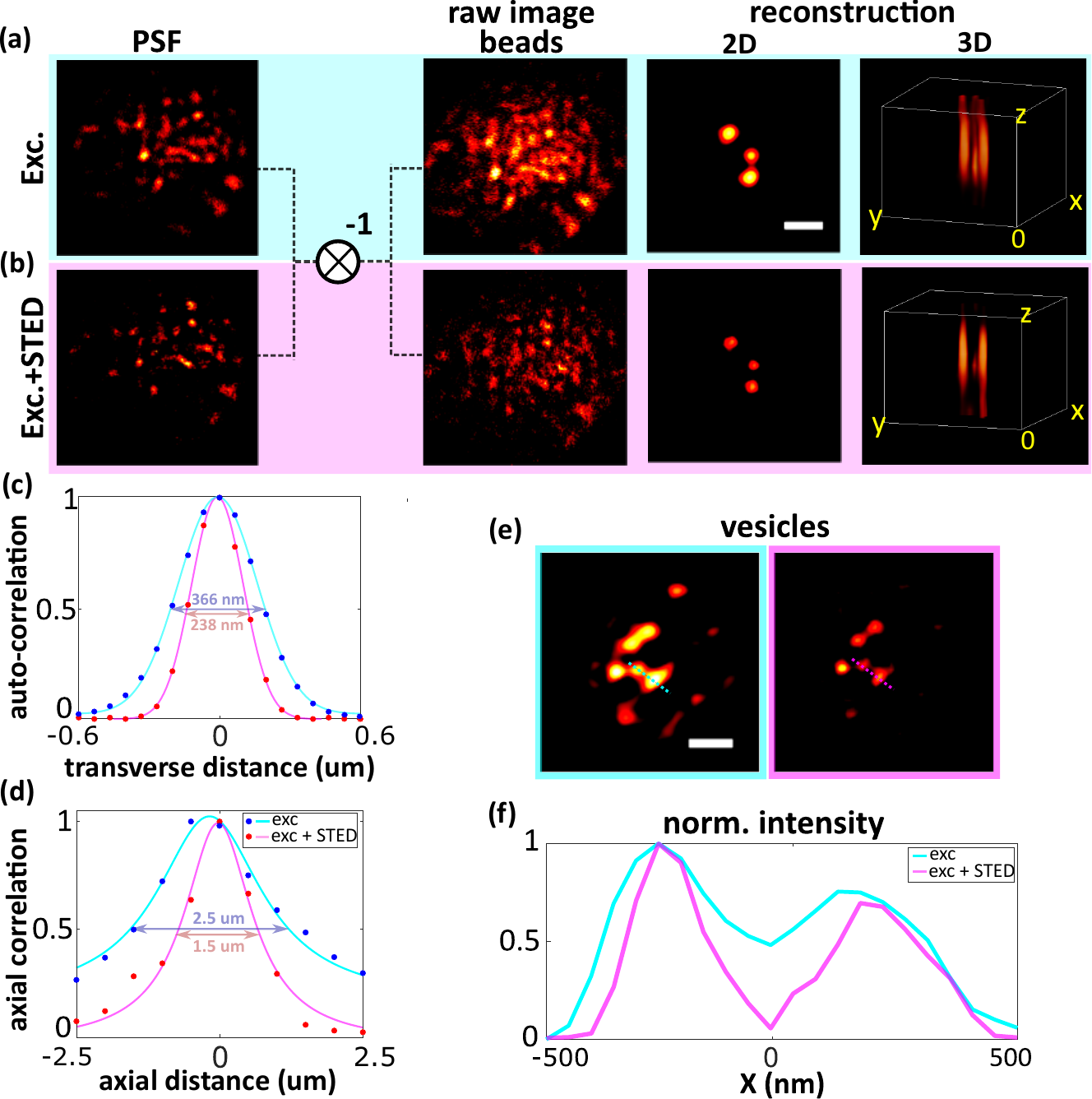}
	\caption{Resolution enhancement using complementary speckle-STED. The PSF, speckle images of beads and reconstruction using FISTA algorithm is shown for our two configurations:excitation and excitation + STED in the first two rows (in cyan and magenta box) respectively (a) and (b). (c) The FWHM of auto-correlation of 2D speckle patterns represented in cyan and blue for excitation and excitation + STED respectively. There is a resolution enhancement of $35\%$. (d) Auto-correlation of 3D speckles (in Z) representing a resolution enhancement in axial direction as well. (e) Reconstructed images of lysosomes in fixed HeLa cells with (f) their line profiles in both configurations.}
	\label{fig:bead-bio}
\end{figure}

 
When the liquid crystal device is tuned so that the excitation and STED beams exhibit a complementary intensity distributions, 3D optical imaging requires the knowledge of the the speckle point spread function (PSF) in 3D. The comparison between the two speckle PSFs obtained with and without the STED beam demonstrate that the depletion laser yields a PSF exhibiting finer structures so effectively giving access to higher spatial frequencies (left images in Fig.~\ref{fig:bead-bio}a and b). The transverse and axial resolution improvement obtained thanks to the contribution of the STED beam are demonstrated by considering the auto-correlation profile of detected fluorescence speckles as shown in Fig.~\ref{fig:bead-bio}c and Fig.~\ref{fig:bead-bio}d, respectively. The resolution enhancement in $xy$ transverse plane (Fig.\ref{fig:bead-bio}(c)) and along the propagation axis are of the order of $\sqrt{2}$  and $1.7$, respectively, for $s\approx4$.\\

Next, samples made of several nano-beads as well as fluorescently labeled vesicles in cell cultures were considered. Typical raw speckle images of beads are shown in Fig.~\ref{fig:bead-bio}a and~\ref{fig:bead-bio}b. Image reconstruction was achieved using a fast-iterative shrinkage thresholding compressive sensing algorithm (FISTA)~\cite{beck2009fast} based on a prior measurement of the 3D speckle PSFs. 
The image of packs of nanobeads retrieved from the saturated depletion condition clearly demonstrates a higher resolution power (Fig.~\ref{fig:bead-bio}b) as compared to the excitation-only condition (Fig.~\ref{fig:bead-bio}a). 
Notably, similar resolution performances are observed between the results achieved using saturated depletion and reconstructed using Wiener deconvolution (Supporting Section 5), although, in theory, iterative algorithms like FISTA have the potential to provide superior resolution by recovering higher spatial frequencies. In our specific experimental conditions, however, Wiener deconvolution of the speckle patterns already recovers a significant portion of the resolution enhancement achieved by the STED process.

As discussed in Ref~\cite{pascucci2019compressive} 3D imaging is possible from 2D speckle images only. Numerical axial sectioning in the detection volume is enabled thanks to the statistical decorrelation of speckle for axial plane separations larger than the axial resolution of the imaging system $\delta z \geq 2n\lambda/{\rm NA}^2$. Under saturated de-excitation conditions, this correlation distance is further reduced (see Fig.~\ref{fig:bead-bio}d), so enabling resolution improvement along the propagation axis. The axial resolution improvement scales as $1/\sqrt{1+s}$~\cite{pascucci2019compressive}. In our case a axial resolution improvement of $\sim 1.7$ is obtained. The three dimensional images of the fluorescent nano-beads are shown on the right images of in Fig.~\ref{fig:bead-bio}a and~\ref{fig:bead-bio}b.
Finally, we demonstrate the applicability of our technique on biological sample (Fig.~\ref{fig:bead-bio}e and f) by imaging lysosomes (with a typical size of $50-500 {\rm nm}$) in fixed cultured HeLa cells where the LAMP1 protein is labeled with Alexa Fluor 647 (ThermoFisher). A clear resolution improvement is observed, consistent with the former speckle characterization we discussed. \par
\section*{Discussion and perspectives}
In this work we implemented a super-resolution STED microscope based on two complementary speckles at two different wavelengths using both fluorescent beads and biological samples. This scheme takes advantage of the intrinsic optical vortices present in random wavefields. Speckle patterns are extensively used as high-contrast structured patterns to enhance the resolution of various imaging techniques. Moreover, the statistical properties of speckle patterns makes them an ideal candidate for compressive sensing schemes, which depending on constraints of the imaging modality, can increase the acquisition rate or the field of view of volumetric imaging, and reduce photobleaching and photo-toxicity. Note that, while photobleaching is comparable to conventional STED due to similar levels of laser exposure, our speckle imaging approach, combined with compressive sensing, provides 3D information from each exposure due to the volumetric distribution of speckle patterns. This is in contrast to traditional point-scanning methods, which yield only 2D information per scan~\cite{marim2009compressed}. We have developed a novel approach employing an SLM and a bi-spectrum q-plate to generate and manipulate the topological properties of speckles at two distinct wavelengths. This method facilitates the creation of both achromatic and complementary speckles with advantageous statistical characteristics for imaging applications. Despite their dependence on interferometric wavelengths, these speckles exhibit an effective correlation range in both modes. By taking advantage of both the speckle orthogonality and optical vortices, we demonstrated the ability to perform super-resolution STED microscopy in 3D from 2D scanned images only. Furthermore, we successfully address the challenge implementing STED configuration where two speckles at from two different laser lines are used for excitation and depletion. Noteworthy, this speckle manipulations at different wavelengths is of interest for other advanced microscopy techniques using more than one laser line such as coherent anti-stokes Raman scattering, stimulated Raman scattering etc~\cite{fantuzzi2023wide}. As a result resolution improvement by a factor $\sqrt{2}$ is obtained. 
We identified that one important factor limiting the resolution enhancement comes from the overall loss of laser power due to the optical efficiency of the speckle generation process and degradation of certain optical elements such as galvanometric mirrors. The resolution improvement in the approach presented in this work results from a statistical average of speckle grain reduction over the whole speckle-PSF surfaces. Therefore, although the complementary pattern approach ensures a high overall correlation between the excitation and the depletion beams, some isolated excitation grains might not achieve the same level of nonlinearity-induced resolution enhancement due to locally diluted STED intensity, so contributing less to the reduction of the effective point spread function.
\section*{Supporting information}
The Supporting Information is available free of charge at \href{https://pubs.acs.org/doi/10.1021/acsphotonics.4c01364}{https://pubs.acs.org/doi/10.1021/acsphotonics.4c01364}, containing 12 pages and 7 figures:\par
Simulated depletion speckle patterns, complementary
speckles vs uniform excitation simulations, optimizing
diffuser parameters $\delta$, optimizing diffuser parameters w,
complete scheme of the speckle-STED setup, fluorescence
depletion using achromatic speckle patterns, and
image reconstruction with Wiener deconvolution
	
\section*{Author contributions statement}
P.A. built up the optical setups and conducted the experiments and data processing. S.C.T., E.B. and H.L. contributed to experiments and data processing. H.Y. built up the bi-spectrum q-plate that E.B. designed and prepared. M.G. conceived the project designed the experiment and supervised the project. P.A., E.B. and M.G. wrote the manuscript, all authors reviewed the manuscript.
\section*{Acknowledgements}
The authors acknowledge Cécile Jouffrey for preparing the HeLa samples, M.G. and P.A. thank Emmanuel Bossy and Sylvain Gigan for stimulating discussions about complementary speckles and Emmanuel Beaurepaire for comments on the manuscript. 
\section*{Funding}
This work was funded by the French Research National Agency (project SpeckleSTED ANR-18-CE42-0008-01). M.G. acknowledges support from Institut Universitaire de France.
\newpage



\begin{thebibliography}{10}
\urlstyle{rm}
\expandafter\ifx\csname url\endcsname\relax
  \def\url#1{\texttt{#1}}\fi
\expandafter\ifx\csname urlprefix\endcsname\relax\def\urlprefix{URL }\fi
\expandafter\ifx\csname doiprefix\endcsname\relax\def\doiprefix{DOI: }\fi
\providecommand{\bibinfo}[2]{#2}
\providecommand{\eprint}[2][]{\url{#2}}

\bibitem{betzig2006imaging}
\bibinfo{author}{Betzig, E.} \emph{et~al.}
\newblock \bibinfo{journal}{\bibinfo{title}{Imaging intracellular fluorescent
  proteins at nanometer resolution}}.
\newblock {\emph{\JournalTitle{science}}} \textbf{\bibinfo{volume}{313}},
  \bibinfo{pages}{1642--1645} (\bibinfo{year}{2006}).

\bibitem{rust2006stochastic}
\bibinfo{author}{Rust, M.~J.}, \bibinfo{author}{Bates, M.} \&
  \bibinfo{author}{Zhuang, X.}
\newblock \bibinfo{journal}{\bibinfo{title}{Stochastic optical reconstruction
  microscopy (storm) provides sub-diffraction-limit image resolution}}.
\newblock {\emph{\JournalTitle{Nature methods}}} \textbf{\bibinfo{volume}{3}},
  \bibinfo{pages}{793} (\bibinfo{year}{2006}).

\bibitem{gustafsson2005nonlinear}
\bibinfo{author}{Gustafsson, M.~G.}
\newblock \bibinfo{journal}{\bibinfo{title}{Nonlinear structured-illumination
  microscopy: wide-field fluorescence imaging with theoretically unlimited
  resolution}}.
\newblock {\emph{\JournalTitle{Proceedings of the National Academy of
  Sciences}}} \textbf{\bibinfo{volume}{102}}, \bibinfo{pages}{13081--13086}
  (\bibinfo{year}{2005}).

\bibitem{hell1994breaking}
\bibinfo{author}{Hell, S.~W.} \& \bibinfo{author}{Wichmann, J.}
\newblock \bibinfo{journal}{\bibinfo{title}{Breaking the diffraction resolution
  limit by stimulated emission: stimulated-emission-depletion fluorescence
  microscopy}}.
\newblock {\emph{\JournalTitle{Optics letters}}} \textbf{\bibinfo{volume}{19}},
  \bibinfo{pages}{780--782} (\bibinfo{year}{1994}).

\bibitem{klar2000fluorescence}
\bibinfo{author}{Klar, T.~A.}, \bibinfo{author}{Jakobs, S.},
  \bibinfo{author}{Dyba, M.}, \bibinfo{author}{Egner, A.} \&
  \bibinfo{author}{Hell, S.~W.}
\newblock \bibinfo{journal}{\bibinfo{title}{Fluorescence microscopy with
  diffraction resolution barrier broken by stimulated emission}}.
\newblock {\emph{\JournalTitle{Proceedings of the National Academy of
  Sciences}}} \textbf{\bibinfo{volume}{97}}, \bibinfo{pages}{8206--8210}
  (\bibinfo{year}{2000}).

\bibitem{movement2001video}
\bibinfo{author}{Movement, S.~V.}
\newblock \bibinfo{journal}{\bibinfo{title}{Video-rate far-field optical
  nanoscopy dissects}}.
\newblock {\emph{\JournalTitle{Chem. Biol}}} \textbf{\bibinfo{volume}{8}},
  \bibinfo{pages}{725} (\bibinfo{year}{2001}).

\bibitem{yang2014large}
\bibinfo{author}{Yang, B.}, \bibinfo{author}{Przybilla, F.},
  \bibinfo{author}{Mestre, M.}, \bibinfo{author}{Trebbia, J.-B.} \&
  \bibinfo{author}{Lounis, B.}
\newblock \bibinfo{journal}{\bibinfo{title}{Large parallelization of sted
  nanoscopy using optical lattices}}.
\newblock {\emph{\JournalTitle{Optics express}}} \textbf{\bibinfo{volume}{22}},
  \bibinfo{pages}{5581--5589} (\bibinfo{year}{2014}).

\bibitem{bergermann20152000}
\bibinfo{author}{Bergermann, F.}, \bibinfo{author}{Alber, L.},
  \bibinfo{author}{Sahl, S.~J.}, \bibinfo{author}{Engelhardt, J.} \&
  \bibinfo{author}{Hell, S.~W.}
\newblock \bibinfo{journal}{\bibinfo{title}{2000-fold parallelized dual-color
  sted fluorescence nanoscopy}}.
\newblock {\emph{\JournalTitle{Optics express}}} \textbf{\bibinfo{volume}{23}},
  \bibinfo{pages}{211--223} (\bibinfo{year}{2015}).

\bibitem{westphal2008video}
\bibinfo{author}{Westphal, V.} \emph{et~al.}
\newblock \bibinfo{journal}{\bibinfo{title}{Video-rate far-field optical
  nanoscopy dissects synaptic vesicle movement}}.
\newblock {\emph{\JournalTitle{Science}}} \textbf{\bibinfo{volume}{320}},
  \bibinfo{pages}{246--249} (\bibinfo{year}{2008}).

\bibitem{schneider2015ultrafast}
\bibinfo{author}{Schneider, J.} \emph{et~al.}
\newblock \bibinfo{journal}{\bibinfo{title}{Ultrafast, temporally stochastic
  sted nanoscopy of millisecond dynamics}}.
\newblock {\emph{\JournalTitle{Nature methods}}} \textbf{\bibinfo{volume}{12}},
  \bibinfo{pages}{827--830} (\bibinfo{year}{2015}).

\bibitem{ding2009supraresolution}
\bibinfo{author}{Ding, J.~B.}, \bibinfo{author}{Takasaki, K.~T.} \&
  \bibinfo{author}{Sabatini, B.~L.}
\newblock \bibinfo{journal}{\bibinfo{title}{Supraresolution imaging in brain
  slices using stimulated-emission depletion two-photon laser scanning
  microscopy}}.
\newblock {\emph{\JournalTitle{Neuron}}} \textbf{\bibinfo{volume}{63}},
  \bibinfo{pages}{429--437} (\bibinfo{year}{2009}).

\bibitem{berning2012nanoscopy}
\bibinfo{author}{Berning, S.}, \bibinfo{author}{Willig, K.~I.},
  \bibinfo{author}{Steffens, H.}, \bibinfo{author}{Dibaj, P.} \&
  \bibinfo{author}{Hell, S.~W.}
\newblock \bibinfo{journal}{\bibinfo{title}{Nanoscopy in a living mouse
  brain}}.
\newblock {\emph{\JournalTitle{Science}}} \textbf{\bibinfo{volume}{335}},
  \bibinfo{pages}{551--551} (\bibinfo{year}{2012}).

\bibitem{oracz2017photobleaching}
\bibinfo{author}{Oracz, J.}, \bibinfo{author}{Westphal, V.},
  \bibinfo{author}{Radzewicz, C.}, \bibinfo{author}{Sahl, S.~J.} \&
  \bibinfo{author}{Hell, S.~W.}
\newblock \bibinfo{journal}{\bibinfo{title}{Photobleaching in sted nanoscopy
  and its dependence on the photon flux applied for reversible silencing of the
  fluorophore}}.
\newblock {\emph{\JournalTitle{Scientific reports}}}
  \textbf{\bibinfo{volume}{7}}, \bibinfo{pages}{11354} (\bibinfo{year}{2017}).

\bibitem{danzl2016coordinate}
\bibinfo{author}{Danzl, J.~G.} \emph{et~al.}
\newblock \bibinfo{journal}{\bibinfo{title}{Coordinate-targeted fluorescence
  nanoscopy with multiple off states}}.
\newblock {\emph{\JournalTitle{Nature Photonics}}}
  \textbf{\bibinfo{volume}{10}}, \bibinfo{pages}{122--128}
  (\bibinfo{year}{2016}).

\bibitem{gottfert2017strong}
\bibinfo{author}{G{\"o}ttfert, F.} \emph{et~al.}
\newblock \bibinfo{journal}{\bibinfo{title}{Strong signal increase in sted
  fluorescence microscopy by imaging regions of subdiffraction extent}}.
\newblock {\emph{\JournalTitle{Proceedings of the National Academy of
  Sciences}}} \textbf{\bibinfo{volume}{114}}, \bibinfo{pages}{2125--2130}
  (\bibinfo{year}{2017}).

\bibitem{staudt2011far}
\bibinfo{author}{Staudt, T.} \emph{et~al.}
\newblock \bibinfo{journal}{\bibinfo{title}{Far-field optical nanoscopy with
  reduced number of state transition cycles}}.
\newblock {\emph{\JournalTitle{Optics express}}} \textbf{\bibinfo{volume}{19}},
  \bibinfo{pages}{5644--5657} (\bibinfo{year}{2011}).

\bibitem{balzarotti2017nanometer}
\bibinfo{author}{Balzarotti, F.} \emph{et~al.}
\newblock \bibinfo{journal}{\bibinfo{title}{Nanometer resolution imaging and
  tracking of fluorescent molecules with minimal photon fluxes}}.
\newblock {\emph{\JournalTitle{Science}}} \textbf{\bibinfo{volume}{355}},
  \bibinfo{pages}{606--612} (\bibinfo{year}{2017}).

\bibitem{weber2021minsted}
\bibinfo{author}{Weber, M.} \emph{et~al.}
\newblock \bibinfo{journal}{\bibinfo{title}{Minsted fluorescence localization
  and nanoscopy}}.
\newblock {\emph{\JournalTitle{Nature photonics}}}
  \textbf{\bibinfo{volume}{15}}, \bibinfo{pages}{361--366}
  (\bibinfo{year}{2021}).

\bibitem{heine2017adaptive}
\bibinfo{author}{Heine, J.} \emph{et~al.}
\newblock \bibinfo{journal}{\bibinfo{title}{Adaptive-illumination sted
  nanoscopy}}.
\newblock {\emph{\JournalTitle{Proceedings of the National Academy of
  Sciences}}} \textbf{\bibinfo{volume}{114}}, \bibinfo{pages}{9797--9802}
  (\bibinfo{year}{2017}).

\bibitem{dreier2019smart}
\bibinfo{author}{Dreier, J.} \emph{et~al.}
\newblock \bibinfo{journal}{\bibinfo{title}{Smart scanning for low-illumination
  and fast resolft nanoscopy in vivo}}.
\newblock {\emph{\JournalTitle{Nature communications}}}
  \textbf{\bibinfo{volume}{10}}, \bibinfo{pages}{556} (\bibinfo{year}{2019}).

\bibitem{alvelid2022event}
\bibinfo{author}{Alvelid, J.}, \bibinfo{author}{Damenti, M.},
  \bibinfo{author}{Sgattoni, C.} \& \bibinfo{author}{Testa, I.}
\newblock \bibinfo{journal}{\bibinfo{title}{Event-triggered sted imaging}}.
\newblock {\emph{\JournalTitle{Nature Methods}}} \textbf{\bibinfo{volume}{19}},
  \bibinfo{pages}{1268--1275} (\bibinfo{year}{2022}).

\bibitem{gustafsson2000surpassing}
\bibinfo{author}{Gustafsson, M.~G.}
\newblock \bibinfo{journal}{\bibinfo{title}{Surpassing the lateral resolution
  limit by a factor of two using structured illumination microscopy}}.
\newblock {\emph{\JournalTitle{Journal of microscopy}}}
  \textbf{\bibinfo{volume}{198}}, \bibinfo{pages}{82--87}
  (\bibinfo{year}{2000}).

\bibitem{chakrova2015studying}
\bibinfo{author}{Chakrova, N.}, \bibinfo{author}{Heintzmann, R.},
  \bibinfo{author}{Rieger, B.} \& \bibinfo{author}{Stallinga, S.}
\newblock \bibinfo{journal}{\bibinfo{title}{Studying different illumination
  patterns for resolution improvement in fluorescence microscopy}}.
\newblock {\emph{\JournalTitle{Optics express}}} \textbf{\bibinfo{volume}{23}},
  \bibinfo{pages}{31367--31383} (\bibinfo{year}{2015}).

\bibitem{mudry2012structured}
\bibinfo{author}{Mudry, E.} \emph{et~al.}
\newblock \bibinfo{journal}{\bibinfo{title}{Structured illumination microscopy
  using unknown speckle patterns}}.
\newblock {\emph{\JournalTitle{Nature Photonics}}}
  \textbf{\bibinfo{volume}{6}}, \bibinfo{pages}{312--315}
  (\bibinfo{year}{2012}).

\bibitem{yilmaz2015speckle}
\bibinfo{author}{Yilmaz, H.} \emph{et~al.}
\newblock \bibinfo{journal}{\bibinfo{title}{Speckle correlation resolution
  enhancement of wide-field fluorescence imaging}}.
\newblock {\emph{\JournalTitle{Optica}}} \textbf{\bibinfo{volume}{2}},
  \bibinfo{pages}{424--429} (\bibinfo{year}{2015}).

\bibitem{yeh2017structured}
\bibinfo{author}{Yeh, L.-H.}, \bibinfo{author}{Tian, L.} \&
  \bibinfo{author}{Waller, L.}
\newblock \bibinfo{journal}{\bibinfo{title}{Structured illumination microscopy
  with unknown patterns and a statistical prior}}.
\newblock {\emph{\JournalTitle{Biomedical optics express}}}
  \textbf{\bibinfo{volume}{8}}, \bibinfo{pages}{695--711}
  (\bibinfo{year}{2017}).

\bibitem{fujita2007high}
\bibinfo{author}{Fujita, K.}, \bibinfo{author}{Kobayashi, M.},
  \bibinfo{author}{Kawano, S.}, \bibinfo{author}{Yamanaka, M.} \&
  \bibinfo{author}{Kawata, S.}
\newblock \bibinfo{journal}{\bibinfo{title}{High-resolution confocal microscopy
  by saturated excitation of fluorescence}}.
\newblock {\emph{\JournalTitle{Physical review letters}}}
  \textbf{\bibinfo{volume}{99}}, \bibinfo{pages}{228105}
  (\bibinfo{year}{2007}).

\bibitem{pascucci2016superresolution}
\bibinfo{author}{Pascucci, M.}, \bibinfo{author}{Tessier, G.},
  \bibinfo{author}{Emiliani, V.} \& \bibinfo{author}{Guillon, M.}
\newblock \bibinfo{journal}{\bibinfo{title}{Superresolution imaging of optical
  vortices in a speckle pattern}}.
\newblock {\emph{\JournalTitle{Physical review letters}}}
  \textbf{\bibinfo{volume}{116}}, \bibinfo{pages}{093904}
  (\bibinfo{year}{2016}).

\bibitem{bender2021circumventing}
\bibinfo{author}{Bender, N.}, \bibinfo{author}{Sun, M.},
  \bibinfo{author}{Y{\i}lmaz, H.}, \bibinfo{author}{Bewersdorf, J.} \&
  \bibinfo{author}{Cao, H.}
\newblock \bibinfo{journal}{\bibinfo{title}{Circumventing the optical
  diffraction limit with customized speckles}}.
\newblock {\emph{\JournalTitle{Optica}}} \textbf{\bibinfo{volume}{8}},
  \bibinfo{pages}{122--129} (\bibinfo{year}{2021}).

\bibitem{pascucci2019compressive}
\bibinfo{author}{Pascucci, M.} \emph{et~al.}
\newblock \bibinfo{journal}{\bibinfo{title}{Compressive three-dimensional
  super-resolution microscopy with speckle-saturated fluorescence excitation}}.
\newblock {\emph{\JournalTitle{Nature communications}}}
  \textbf{\bibinfo{volume}{10}}, \bibinfo{pages}{1327} (\bibinfo{year}{2019}).

\bibitem{lim2008wide}
\bibinfo{author}{Lim, D.}, \bibinfo{author}{Chu, K.~K.} \&
  \bibinfo{author}{Mertz, J.}
\newblock \bibinfo{journal}{\bibinfo{title}{Wide-field fluorescence sectioning
  with hybrid speckle and uniform-illumination microscopy}}.
\newblock {\emph{\JournalTitle{Optics letters}}} \textbf{\bibinfo{volume}{33}},
  \bibinfo{pages}{1819--1821} (\bibinfo{year}{2008}).

\bibitem{donoho2006compressed}
\bibinfo{author}{Donoho, D.~L.}
\newblock \bibinfo{journal}{\bibinfo{title}{Compressed sensing}}.
\newblock {\emph{\JournalTitle{IEEE Transactions on information theory}}}
  \textbf{\bibinfo{volume}{52}}, \bibinfo{pages}{1289--1306}
  (\bibinfo{year}{2006}).

\bibitem{bernet2011lensless}
\bibinfo{author}{Bernet, S.}, \bibinfo{author}{Harm, W.},
  \bibinfo{author}{Jesacher, A.} \& \bibinfo{author}{Ritsch-Marte, M.}
\newblock \bibinfo{journal}{\bibinfo{title}{Lensless digital holography with
  diffuse illumination through a pseudo-random phase mask}}.
\newblock {\emph{\JournalTitle{Optics express}}} \textbf{\bibinfo{volume}{19}},
  \bibinfo{pages}{25113--25124} (\bibinfo{year}{2011}).

\bibitem{antipa2018diffusercam}
\bibinfo{author}{Antipa, N.} \emph{et~al.}
\newblock \bibinfo{journal}{\bibinfo{title}{Diffusercam: lensless
  single-exposure 3d imaging}}.
\newblock {\emph{\JournalTitle{Optica}}} \textbf{\bibinfo{volume}{5}},
  \bibinfo{pages}{1--9} (\bibinfo{year}{2018}).

\bibitem{nye1974dislocations}
\bibinfo{author}{Nye, J.~F.} \& \bibinfo{author}{Berry, M.~V.}
\newblock \bibinfo{journal}{\bibinfo{title}{Dislocations in wave trains}}.
\newblock {\emph{\JournalTitle{Proceedings of the Royal Society of London. A.
  Mathematical and Physical Sciences}}} \textbf{\bibinfo{volume}{336}},
  \bibinfo{pages}{165--190} (\bibinfo{year}{1974}).

\bibitem{bender2018customizing}
\bibinfo{author}{Bender, N.}, \bibinfo{author}{Y{\i}lmaz, H.},
  \bibinfo{author}{Bromberg, Y.} \& \bibinfo{author}{Cao, H.}
\newblock \bibinfo{journal}{\bibinfo{title}{Customizing speckle intensity
  statistics}}.
\newblock {\emph{\JournalTitle{Optica}}} \textbf{\bibinfo{volume}{5}},
  \bibinfo{pages}{595--600} (\bibinfo{year}{2018}).

\bibitem{chmyrov2013nanoscopy}
\bibinfo{author}{Chmyrov, A.} \emph{et~al.}
\newblock \bibinfo{journal}{\bibinfo{title}{Nanoscopy with more than
  100,000'doughnuts'}}.
\newblock {\emph{\JournalTitle{Nature methods}}} \textbf{\bibinfo{volume}{10}},
  \bibinfo{pages}{737--740} (\bibinfo{year}{2013}).

\bibitem{gateau2017complementary}
\bibinfo{author}{Gateau, J.}, \bibinfo{author}{Rigneault, H.} \&
  \bibinfo{author}{Guillon, M.}
\newblock \bibinfo{journal}{\bibinfo{title}{Complementary speckle patterns:
  deterministic interchange of intrinsic vortices and maxima through scattering
  media}}.
\newblock {\emph{\JournalTitle{Physical Review Letters}}}
  \textbf{\bibinfo{volume}{118}}, \bibinfo{pages}{043903}
  (\bibinfo{year}{2017}).

\bibitem{gateau2019topological}
\bibinfo{author}{Gateau, J.}, \bibinfo{author}{Claude, F.},
  \bibinfo{author}{Tessier, G.} \& \bibinfo{author}{Guillon, M.}
\newblock \bibinfo{journal}{\bibinfo{title}{Topological transformations of
  speckles}}.
\newblock {\emph{\JournalTitle{Optica}}} \textbf{\bibinfo{volume}{6}},
  \bibinfo{pages}{914--920} (\bibinfo{year}{2019}).

\bibitem{bianchini2012single}
\bibinfo{author}{Bianchini, P.}, \bibinfo{author}{Harke, B.},
  \bibinfo{author}{Galiani, S.}, \bibinfo{author}{Vicidomini, G.} \&
  \bibinfo{author}{Diaspro, A.}
\newblock \bibinfo{journal}{\bibinfo{title}{Single-wavelength two-photon
  excitation--stimulated emission depletion (sw2pe-sted) superresolution
  imaging}}.
\newblock {\emph{\JournalTitle{Proceedings of the National Academy of
  Sciences}}} \textbf{\bibinfo{volume}{109}}, \bibinfo{pages}{6390--6393}
  (\bibinfo{year}{2012}).

\bibitem{vesga2019focusing}
\bibinfo{author}{Vesga, A.~G.} \emph{et~al.}
\newblock \bibinfo{journal}{\bibinfo{title}{Focusing large spectral bandwidths
  through scattering media}}.
\newblock {\emph{\JournalTitle{Optics express}}} \textbf{\bibinfo{volume}{27}},
  \bibinfo{pages}{28384--28394} (\bibinfo{year}{2019}).

\bibitem{zhu2020chromato}
\bibinfo{author}{Zhu, L.} \emph{et~al.}
\newblock \bibinfo{journal}{\bibinfo{title}{Chromato-axial memory effect
  through a forward-scattering slab}}.
\newblock {\emph{\JournalTitle{Optica}}} \textbf{\bibinfo{volume}{7}},
  \bibinfo{pages}{338--345} (\bibinfo{year}{2020}).

\bibitem{arjmand2021three}
\bibinfo{author}{Arjmand, P.}, \bibinfo{author}{Katz, O.},
  \bibinfo{author}{Gigan, S.} \& \bibinfo{author}{Guillon, M.}
\newblock \bibinfo{journal}{\bibinfo{title}{Three-dimensional broadband light
  beam manipulation in forward scattering samples}}.
\newblock {\emph{\JournalTitle{Optics Express}}} \textbf{\bibinfo{volume}{29}},
  \bibinfo{pages}{6563--6581} (\bibinfo{year}{2021}).

\bibitem{yan2015q}
\bibinfo{author}{Yan, L.} \emph{et~al.}
\newblock \bibinfo{journal}{\bibinfo{title}{Q-plate enabled spectrally diverse
  orbital-angular-momentum conversion for stimulated emission depletion
  microscopy}}.
\newblock {\emph{\JournalTitle{Optica}}} \textbf{\bibinfo{volume}{2}},
  \bibinfo{pages}{900--903} (\bibinfo{year}{2015}).

\bibitem{guillon2017vortex}
\bibinfo{author}{Guillon, M.} \emph{et~al.}
\newblock \bibinfo{journal}{\bibinfo{title}{Vortex-free phase profiles for
  uniform patterning with computer-generated holography}}.
\newblock {\emph{\JournalTitle{Optics Express}}} \textbf{\bibinfo{volume}{25}},
  \bibinfo{pages}{12640--12652} (\bibinfo{year}{2017}).

\bibitem{piccirillo2010photon}
\bibinfo{author}{Piccirillo, B.}, \bibinfo{author}{D’Ambrosio, V.},
  \bibinfo{author}{Slussarenko, S.}, \bibinfo{author}{Marrucci, L.} \&
  \bibinfo{author}{Santamato, E.}
\newblock \bibinfo{journal}{\bibinfo{title}{Photon spin-to-orbital angular
  momentum conversion via an electrically tunable q-plate}}.
\newblock {\emph{\JournalTitle{Applied Physics Letters}}}
  \textbf{\bibinfo{volume}{97}} (\bibinfo{year}{2010}).

\bibitem{tkachenko2006nematic}
\bibinfo{author}{Tkachenko, V.} \emph{et~al.}
\newblock \bibinfo{journal}{\bibinfo{title}{Nematic liquid crystal optical
  dispersion in the visible-near infrared range}}.
\newblock {\emph{\JournalTitle{Molecular Crystals and Liquid Crystals}}}
  \textbf{\bibinfo{volume}{454}}, \bibinfo{pages}{263--665}
  (\bibinfo{year}{2006}).

\bibitem{freund19981001}
\bibinfo{author}{Freund, I.}
\newblock \bibinfo{journal}{\bibinfo{title}{1001'correlations in random wave
  fields}}.
\newblock {\emph{\JournalTitle{Waves in random media}}}
  \textbf{\bibinfo{volume}{8}}, \bibinfo{pages}{119} (\bibinfo{year}{1998}).

\bibitem{beck2009fast}
\bibinfo{author}{Beck, A.} \& \bibinfo{author}{Teboulle, M.}
\newblock \bibinfo{journal}{\bibinfo{title}{A fast iterative
  shrinkage-thresholding algorithm for linear inverse problems}}.
\newblock {\emph{\JournalTitle{SIAM journal on imaging sciences}}}
  \textbf{\bibinfo{volume}{2}}, \bibinfo{pages}{183--202}
  (\bibinfo{year}{2009}).

\bibitem{marim2009compressed}
\bibinfo{author}{Marim, M.~M.}, \bibinfo{author}{Angelini, E.~D.} \&
  \bibinfo{author}{Olivo-Marin, J.-C.}
\newblock \bibinfo{title}{A compressed sensing approach for biological
  microscopic image processing}.
\newblock In \emph{\bibinfo{booktitle}{2009 IEEE International Symposium on
  Biomedical Imaging: From Nano to Macro}}, \bibinfo{pages}{1374--1377}
  (\bibinfo{organization}{IEEE}, \bibinfo{year}{2009}).

\bibitem{fantuzzi2023wide}
\bibinfo{author}{Fantuzzi, E.~M.} \emph{et~al.}
\newblock \bibinfo{journal}{\bibinfo{title}{Wide-field coherent anti-stokes
  raman scattering microscopy using random illuminations}}.
\newblock {\emph{\JournalTitle{Nature Photonics}}}
  \textbf{\bibinfo{volume}{17}}, \bibinfo{pages}{1097--1104}
  (\bibinfo{year}{2023}).

\end{thebibliography}

\end{document}